\documentclass[aps,twocolumn,superscriptaddress]{revtex4-1}
\usepackage{color}
\usepackage{graphicx}
\usepackage{epstopdf}
\usepackage{amsmath}
\usepackage{mathtools}
\usepackage[colorlinks,linkcolor=blue,anchorcolor=blue,citecolor=blue,urlcolor=blue]{hyperref}
\newcommand{\bea}{\begin{eqnarray}}
\newcommand{\eea}{\end{eqnarray}}

\newcommand{\bA}{\mathbf{A}}

\newcommand{\bn}{\mathbf{n}}

\newcommand{\ie}{\textit{i.e.{ }}}
\newcommand{\eg}{\textit{e.g.{ }}}

\newcommand{\etal}{\textit{et al.{ }}}

\newcommand{\me}{\mathrm{e}}

\begin{document}
\title{Critical exponents of nonlinear sigma model on Grassmann manifold $U(N)/U(m)U(N-m)$ by $1/N$ expansion}
\author{Shan-Yue Wang}
\affiliation{National Laboratory of Solid State Microstructures $\&$ School of Physics, Nanjing University, Nanjing, 210093, China}
\author{Da Wang} %\email{dawang@nju.edu.cn}
\affiliation{National Laboratory of Solid State Microstructures $\&$ School of Physics, Nanjing University, Nanjing, 210093, China}
\author{Qiang-Hua Wang} %\email{qhwang@nju.edu.cn}
\affiliation{National Laboratory of Solid State Microstructures $\&$ School of Physics, Nanjing University, Nanjing, 210093, China}
\affiliation{Collaborative Innovation Center of Advanced Microstructures, Nanjing University, Nanjing 210093, China}
\begin{abstract}
  Motivated by the numerical evidence of a continuous phase transition between antiferromagnetic and paramagnetic phases
  in the half-filled SU(N) Hubbbard model, we studied its low energy nonlinear sigma model defined on Grassman manifold
  $U(N)/U(m)U(N-m)$ using the complex projective presentation, which is a direct generalization of the widely studied
  CP$^{N-1}$ model (corresponding to $m=1$). With the $1/N$ expansion technique up to the first order by fixing $m$ in space dimension
  $2<d<4$, we calculate the critical exponents of the Neel moment, which are found to be only functions of $m/N$. Our
  results indicate that larger $m$ effectively reduces $N$ and thus brings stronger fluctuations around the saddle point at $N=\infty$.
\end{abstract}
\maketitle

\section{Introduction}
% SU(N)-spin symmetry breaking: dependence of m physical realization: cold atom fermionic system and QMC results
% universality class, CP model
Spin is a fundamental property of elementary particles inherited from the spatial SO(3) symmetry. For a single electron,
there are only three independent spin directions corresponding to three generators of the SO(3) or SU(2) group. In solids, different
electrons can be combined by Hund's coupling to form larger spins as different representations of the SO(3) or SU(2)
group. If all other degrees of freedom such as charge are frozen, the system is said to be described by an effective
spin model, \eg the famous Heisenberg model. Due to the lacking of small parameters, the spin-$1/2$ Heisenberg model is
difficult to solve by perturbation theories. In such a background, people introduced the SU(N) spin such that $1/N$ can
now be taken as a perturbation parameter.
\cite{affleck1985,affleck1988,*marston1989,arovas1988,read1989,*read1989a,*read1990} One important consequence of the
larger SU(N) group is that it has more ($N^2-1$) independent spin directions (generators), which can be broken into
different kinds of magnetic ordered states. Taking antiferromagnetic (AF) order as an example, the spin in A-sublattice
is represented by a Young tableau with $m$ rows and $n_c$ columns, while the spin in B-sublattice belongs to its
conjugate representation, \ie $N-m$ rows and $n_c$ columns. In recent years, the SU(N) Heisenberg model and its
variations have been extensively studied mainly focusing on the case of $m=1$ (called fundamental representation) from
different numerical approaches.
\cite{harada2003,buchta2007,kawashima2007,arovas2008,beach2009,lou2009,rachel2009,kaul2012,harada2013,nataf2014,okubo2015,suzuki2015,nataf2016}

On the other hand, the fast development of cold atom physics provides an ingenious physical realization of the SU(N)
spin since the carriers in an optical lattice is not electrons but atoms which can have larger hyperfine nuclear spins
which span the internal space.
\cite{wu2003,wu2005,*wu2006,honerkamp2004,taie2010,desalvo2010,krauser2012,zhang2014,hart2015} To mimic such a
fermionic system, people have paid much effort to study the SU(N) Hubbard model.
\cite{honerkamp2004,assaad2005,cai2013a,cai2013,wang2014,zhou2014,*zhou2016,*zhou2017} On the square lattice with
particle hole symmetry, AF order with $m=N/2$ (called self conjugate representation) are found to be suppressed by
either large $N$ or large enough Hubbard interaction and even driven to valence bond solid (VBS) state.
\cite{cai2013,wang2014} In special, very recently, by a high precision QMC calculation on the SU(6) Hubbard model, one of the authors
with his collaborators have found evidence that the AF phase transition is continuous with its critical exponents obtained, 
\cite{wang2014,wang2018} which naturally posts an interesting theoretical question on how to describe such a novel critical point
or universality class. The case of $m=1$ has been widely studied based on its low energy effective theory
called CP$^{N-1}$ model, which is in fact a multi-component generalization of the Ginzburg-Landau theory.
\cite{halperin1974,hikami1979,read1989,read1989a,starykh1994,irkhin1996,kaul2008,block2013,demidio2017} However, the
general case with $m>1$ has received very little attention. \cite{macfarlane1979,hikami1980,duerksen1981,maharana1983}
To the best of our knowledge, particularly, the theoretical calculations of critical exponents belonging to either AF or
VBS are still missing. In this work, we focus on the AF phase transition with general $m$. We first derive a generalized
CP$^{N-1}$ model from the nonlinear sigma model on Grassmann manifold $U(N)/U(m)U(N-m)$, and then use $1/N$ expansion
technique up to the first order by fixing $m$ to calculate the critical exponents. Our results of these critical exponents, remarkably,
show only dependence on $m/N$ indicating that larger $m$ brings stronger fluctuations around the saddle point solutions
in the large $N$ limit.

\section{Model}
Let us start from the fermionic representation (\eg in the SU(N) Hubbard model) of the SU(N) AF order parameter \bea
\mathcal{N}_i^b=(-1)^i\langle \Psi_i^\dagger \sigma_b \Psi_i\rangle \eea where $\Psi_i$ is an N-flavor fermion field and
$\sigma_b$ ($1\le b\le N^2-1$) denote the SU(N) group generators satisfying a normalization condition
$\mathrm{tr}(\sigma_b^2)=N$ within this work. In the normal (paramagnetic) state all these $(N^2-1)$ branches of magnetic modes
degenerate as a result of the SU(N) symmetry. While in the AF state, the SU(N) symmetry can be broken into different
irreducible representations which are characterized by $m$: there are $m$ and $(N-m)$ fermions on two kinds of
sublattice, respectively. For simplicity but without loss of generality, we can always use a diagonal matrix \bea
\sigma_m=\text{diag}\left[P_+,\cdots,P_+,-P_,\cdots,-P_-\right] \eea to represent the ordered moment where
$P_+=P_-^{-1}=\sqrt{(N-m)/m}$, and thus the AF order parameter is simply given by \bea
\mathcal{N}_i^m=(-1)^i\left[P_+\sum_{\alpha=1}^m n_{i\alpha}-P_-\sum_{\alpha=m+1}^{N}n_{i\alpha} \right]. \eea where
$n_{i\alpha}$ is the fermion occupation number. 

Near the AF phase transition of the SU(N) Hubbard model, if fermionic excitations are gapped out (which is seen in QMC
works with $m=N/2$ \cite{cai2013,wang2014}) and thus the critical behavior is governed by the nonlinear sigma model \bea
\label{eq:NLsM}S=\frac{\rho_s}{2}\int \partial_\mu\bn \cdot \partial_\mu\bn \eea where $\rho_s$ is the spin stiffness,
$\mu$ denotes the space indices, $\int$ means spatial integral and $\bn$ is the normalized Neel moment $\mathcal{N}$.
The action $S$ is directly related to the partition function $Z=\int \me^{-S}$. In this work, we parametrize $\bn$ using
the complex representation \bea \label{eq:cprep}n_b=\sum_{i=1}^m z_i^\dag \sigma_b z_i=\mathrm{Tr}(Z^\dag\sigma_b Z)
\eea where $z_i$ is an N-flavor boson field (spinon) and $m$ copies of $z_i$ are put together to form a complex
$N\times m$ matrix $Z$. Next, applying the Fierz identity of the $SU(N)$ group \bea \label{eq:Fierz}
(\sigma_b)_{\alpha\beta}
(\sigma_b)_{\gamma\delta}+\delta_{\alpha\beta}\delta_{\gamma\delta}=N\delta_{\alpha\delta}\delta_{\beta\gamma} \eea we
get \bea \bn\cdot\bn = N\mathrm{Tr}(ZZ^\dag Z Z^\dag)-\mathrm{Tr}(ZZ^\dag)^2 \eea In order to maintain $\bn\cdot\bn=1$,
we impose a constraint condition \bea Z^\dag Z=\frac{1}{\sqrt{m(N-m)}}I \label{eq:normalizeZ}\eea where $I$ is a
rank-$m$ unit matrix. This constraint equation indicates that the spinon field lives on the Grassmann manifold
$U(N)/U(m)U(N-m)$. 

Substituting the complex representation Eq.~\ref{eq:cprep} into Eq.~\ref{eq:NLsM} and again applying Fierz identity
Eq.~\ref{eq:Fierz}, we have \bea S &=& N\rho_s \int\mathrm{Tr}\left[-(i\partial_\mu Z^\dagger Z) (-iZ^\dag \partial_\mu
Z) \right. \nonumber\\  &+& \left. Z^\dag Z (i\partial_\mu Z^\dagger)(-i\partial_\mu Z)\right] .\eea The first term can
be further decoupled by introducing an auxiliary Hubbard-Stratonovich field $\bA$, leading to \bea \label{eq:cpNm0} S=\frac{N\rho_s}{\sqrt{m(N-m)}}\int
\mathrm{Tr}\left[ (i\partial_\mu Z^\dag+A_\mu Z^\dag)(-i\partial_\mu Z+ZA_\mu)  \right] \nonumber\\ \eea Here $A_\mu$ is an $m\times
m$ matrix and thus resembles a non-Abelian gauge field. The $Z$ field can be further rescaled to eliminate the prefactor
$N\rho_s/\sqrt{m(N-m)}$, which after the rescaling appears in the right hand side of the constraint condition Eq.~\ref{eq:normalizeZ}. Using the
rescaled field and introducing a real Lagrangian multiplier matrix $\lambda$ to incorporate the constraint condition, we
arrive at \bea \label{eq:cpNm} S&=&\int \mathrm{Tr}\left[ (i\partial_\mu Z^\dag+A_\mu Z^\dag)(-i\partial_\mu Z+ZA_\mu)
\right] \nonumber \\ &+& i\int \mathrm{Tr}\left\{\lambda\left[Z^\dag Z-\frac{N\rho_s}{m(N-m)}I\right]\right\} .
\eea This model is a direct generalization of the famous CP$^{N-1}$ model (as a special case with $m=1$). The main
difference is: all $Z$, $A$, $\lambda$ fields now become matrices (or multi-components). 
In history, such a model was first proposed forty
years ago by MacFarlane \cite{macfarlane1979} and latter studied by some other researchers for different purposes
\cite{hikami1980,duerksen1981,maharana1983}. However, due to the missing of a real physical system governed by such a
model, it received less and less attentions. In this work, we rediscover it and apply to the SU(N) spin system to study
the critical behaviors. We will only constraint us in the renormalized classical region (keeping only the smallest Matsubara
frequency) since it is sufficient to obtain the critical exponents. \cite{irkhin1996}

\section{Large-N limit}
We first study the large N limit, since in which case the saddle point solution becomes exact as a result of the global
prefactor $N$ after tracing out the $Z$-field. The saddle point condition $\delta S/\delta bA_\mu$ gives $A_\mu=0$ which
means there is no gauge fluctuation. While the other saddle point condition $\delta S/\delta \lambda=0$ just gives the
constraint condition. In order to describe the ordered phase, we assume \bea Z_{\alpha i}=z_0\delta_{\alpha i}+z_{\alpha
i}. \eea where $\alpha\le N$ and $i\le m$. The first term means condensation occurs when $z_0\ne0$ with a remaining
$SU(m)\times SU(N-m)$ symmetry and the second term describes spinon fluctuations. Now the constraint condition becomes
\bea \label{eq:Ninfty} z_0^2 + NT\int_k \frac{1}{k^2} = \frac{N\rho_s}{{m(N-m)}}. \eea 

We first examine the critical exponents of the $Z$ field. At $T=T_c$, $z_0=0$ and $G_z(k)=k^{-2}$ which gives $\eta_z=0$
by definition $G_z(k)\sim k^{\eta-2}$. The exponent $\gamma_z$ is related to the uniform static susceptibility, or
equivalently $G_z(k=0)$ above $T_c$. In this case, a mass term $\Delta$ should be added to retain the constraint
condition, \ie $G_z(k)=(k^2+\Delta)^{-1}$. \cite{ma1973} Then from Eq.~\ref{eq:Ninfty}, we get $\Delta\propto t^{2/(d-2)}$ where
$t=(T-T_c)/T_c$. Comparing with the definition $G_z(k=0)\sim t^{-\gamma_z}$, we obtain $\gamma_z=2/(d-2)$. From $\eta_z$
and $\gamma_z$, all other exponents of the $Z$ fields can be obtained from scaling relations. In particular,
$\nu_z=1/(d-2)$ which should be independent on specified fields and thus we can drop its subscript $z$, \ie
$\nu=1/(d-2)$. 

Next, let us focus on the magnetic moment as a particle-hole pair of the $Z$ field. The order parameter $M$ is
determined by the condensed $Z$ field by definition, \ie $M=mP_+z_0^2$. From Eq.~\ref{eq:Ninfty}, we obtain
$z_0^2\propto t$ and thus $M\propto t$. As a result, the exponent $\beta_n=1$ by definition $M\propto |t|^{\beta_n}$.
Then, together with $\nu=1/(d-2)$ obtained from above, all remaining exponents in the magnetic channel are readily obtained, 
\eg $\eta_n=d-2$ and $\gamma_n=(4-d)/(d-2)$.

\section{1/N expansion}

\begin{figure}
\includegraphics[width=0.4\textwidth]{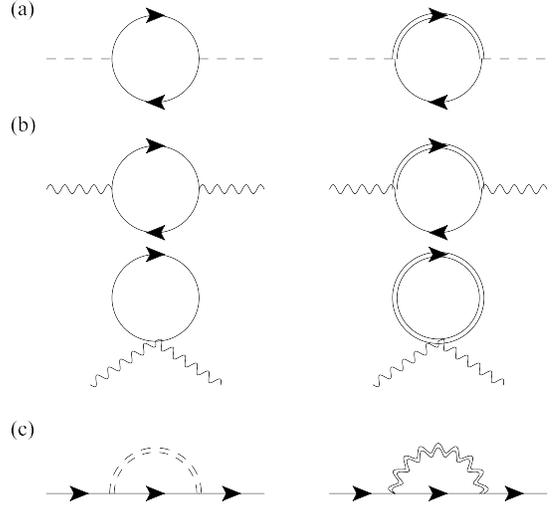}
\caption{\label{fig:self} (a) and (b) show Vacuum polarizations of $i\lambda$ and $A$, respectively. (c) shows the self energy of the $z$ field. Solid (double) lines represent propagators of the (condensed) $Z$ field. Dashed/wavy lines represent undressed $i\lambda$/$A$ fields and their double lines mean propagators.}
\end{figure}

Next, let us go
beyond the saddle point approximation by performing $1/N$ expansion calculation while fixing $m$. Our strategy mainly follows the work by
Irkhin \etal \cite{irkhin1996} which only considered the case with $m=1$. Up to the first order of $1/N$, both
$i\lambda$ and $A$ fields acquire dynamics from their vacuum polarization processes as shown in Fig.~\ref{fig:self}(a) and (b),
\bea \Pi_{ij}^\lambda(q)&=&2z_0^2\frac{1}{q^2} + NT\int_k \frac{1}{k^2(k+q)^2} \nonumber\\
                        &=&\frac{2z_0^2}{q^2}+\frac{TNK_dA_d}{q^{4-d}} \equiv \Pi_\lambda(q) \eea and \bea
\Pi_{ij}^{\mu\nu}(q) &=& 2z_0^2 \frac{q_\mu q_\nu}{q^2} + NT\int_k \frac{(2k_\mu+q_\mu)(2k_\nu+q_\nu)}{k^2(k+q)^2}
\nonumber\\ && - 2z_0^2 \delta_{\mu\nu} -2NT\int_k \frac{1}{k^2}\delta_{\mu\nu}  \nonumber\\
            &=&\left(2z_0^2+\frac{TNK_dA_d}{d-1} q^{d-2}\right)\left(\frac{q_\mu q_\nu}{q^2}-\delta_{\mu\nu}\right)
\nonumber\\ &\equiv& \Pi_A(q)\left(\frac{q_\mu q_\nu}{q^2}-\delta_{\mu\nu}\right) \eea where \bea
K_d^{-1}=2^{d-1}\pi^{d/2}\Gamma\left(\frac{d}{2}\right) \eea and \bea
A_d=\frac{\Gamma(d-2)\Gamma(2-d/2)\Gamma^2(d/2-1)}{2\Gamma(d-2)} \eea One good property is that both
$\Pi_{ij}^\lambda(q)$ and $\Pi_{ij}^{\mu\nu}(q)$ show no dependence on their subscript $ij$ and $m$, which will greatly
simplify the following calculations. From the vacuum polarizations, we get their propagators
$D_\lambda(q)=-\Pi_\lambda^{-1}$ and $D_A^{\mu\nu}(q)=-\Pi_A^{-1}\left(\frac{q_\mu q_\nu}{q^2}-\delta_{\mu\nu}\right)$
under the Landau gauge. For other gauges, the propagator will explicitly depend on the gauge fixing parameter $\xi$. 
However, since the magnetic moment is gauge independent, the results of its critical exponents should be independent of $\xi$. \cite{kaul2008}
%{\color{red} @WSY: if you like, you can add the gauge fixing parameter $\xi$ explicitly.}

\begin{figure}
\includegraphics[width=0.45\textwidth]{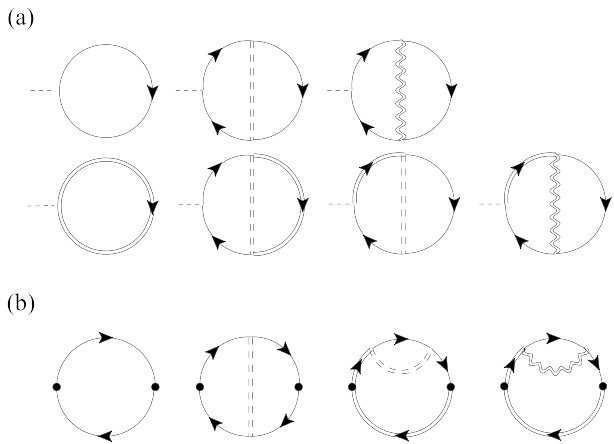}
\caption{\label{fig:suscept} (a) gives the constraint condition: summation of these diagrams equals $N\rho_s/m(N-m)$. (b) shows the uniform susceptibility of $\bn$, in which the external vertex is $\sigma_m$. }
\end{figure}

The self energy of the $Z$ field shown in Fig.~\ref{fig:self}(c) is $\Sigma=\Sigma_\lambda+\Sigma_A$ where \bea
\Sigma_\lambda(k)=-mT\int_q \left[\frac{1}{(k+q)^2}-\frac{1}{q^2}\right]\Pi_\lambda^{-1}(q) \eea and \bea \Sigma_A(k)
&=& mT\int_q \frac{1}{(k+q)^2} D_A^{\mu\nu}(q) (2k_\mu+q_\mu) (2k_\nu+q_\nu) \nonumber \\&=& 4mT\int_q
\frac{1}{(k+q)^2}\Pi_A^{-1}\left[ k^2-\frac{(k\cdot q)^2}{q^2} \right] \eea After completing the bubble integrals, and
extracting the coefficient of the $k^2\ln k$ term, we immediately get $\eta_z=-\frac{(4d-7)m}{NA_d}$, which was first obtained by
Hikami \cite{hikami1980}. However, his calculation was based on single particle propagators and thus can only gave the
critical exponents of the $Z$ field. On the other hand, here, we are more interested in the critical exponents of the
physical field $\bn$. To this end, we have to consider the correlation functions of $\bn$, which is a two-particle
propagator. Keeping only the uniform term ($q=0$) as shown in Fig.~\ref{fig:suscept}(b), the susceptibility of $\bn$ is
\bea \chi_0=m^2P_+^2 z_0^4\left[ 1-  2mT\int_k \frac{\Pi_\lambda^{-1}}{k^4} + 4
\left.\frac{\Sigma(k)}{k^2}\right|_{k\rightarrow0} \right] \eea which gives the magnetic moment square $M^2$ as a
function $z_0$. In order to get $\beta_n$, we still need the relation between $z_0$ and $t$, which can be obtained
from the constraint condition up to order of $1/N$ as shown in Fig.~\ref{fig:suscept}(a), \bea &&z_0^2 \left[ 1-
mT\int_k \frac{\Pi_\lambda^{-1}}{k^4} + 2 \left.\frac{\Sigma(k)}{k^2}\right|_{k\rightarrow0}\right] + NT\int_k
\frac{1}{k^2} \nonumber\\&&+ NT\int_k \frac{\Sigma(k)}{k^4} = \frac{N\rho_s}{m(N-m)}. \eea After some lengthy
but straightforward algebra, we get \bea \beta_n=1-\frac{2m(d^2-d+2)}{NA_d}. \eea Taking together with the
correlation length exponent $\nu=\nu_z$ which had already obtained in Ref.~\cite{hikami1980} \bea
\nu=\frac{1}{d-2}\left[ 1-\frac{2md(d-1)}{NA_d} \right], \eea we get the remaining critical exponents, \bea
\eta_n=(d-2)\left[1-\frac{8m}{NA_d}\right] \eea and \bea \gamma_n=\frac{4-d}{d-2}\left[
1+\frac{8m(d-2)}{(4-d)NA_d}-\frac{2md(d-1)}{NA_d} \right]. \eea Although the calculation is somewhat tedious, the
result is quite clear: the critical exponents only depend on $m/N$ rather than $1/N$ as comparing with the
CP$^{N-1}$ model \cite{irkhin1996}. Hence, up to the first order of $1/N$, the critical exponents of $m>1$ are just
equal to those obtained from the CP$^{(N/m)-1}$ model. 

\section{Discussion}
In summary, we have derived a generalized CP$^{N-1}$ model on Grassman manifold $U(N)/U(m)U(N-m)$ and performed a $1/N$
expansion calculation of the critical exponents for $2<d<4$, which show dependence on $m/N$. Unfortunately, applying our
results to the AF phase transition with self conjugate representation $m=N/2$, we get negative critical exponents which
is inconsistent with the QMC simulations on SU(6) Hubbard model. \cite{wang2014,wang2018} Several remarks are given: (1)Larger $m$ may spoil the
first order perturbation and higher order calculations are in need; (2)The generalized CP model is not sufficient to
describe the AF phase transition since the paramagnetic phase maybe a VBS and a deconfined theory with an additional
topological term may be necessary to get the correct critical exponents. \cite{senthil2004,motrunich2004} 
Furthermore, for a deconfined QCP, there may be two
length scales and thus two $\nu$. \cite{Shao2016} (3)Our calculation is for a classical (finite temperature) phase
transition in dimension $d$, which may have different critical exponents from the quantum phase transition in dimension
$d-1$ as a result of the dynamical exponent $z$. \cite{sondhi1997}

\section{acknowledgement}
DW thanks A. A. Katanin, V. Y. Irkhin, and C. Wu for valuable discussions. In particular, DW thanks A. A. Katanin for his generous help on the calculation details in Ref.~\onlinecite{irkhin1996} and thanks C. Wu for his help on the SU(N) symmetry analysis. This work is supported by NSFC (Nos. 11504164 and 11574134).

{\it Note added.} After completing our manuscript, we became aware of another related work \cite{das2018} on a similar model from the RG approach, which, however, only obtained single particle critical exponents.

\bibliography{cpNm}
\end{document}